\documentstyle[fleqn]{tp}

\input psfig

\def\etal{{\it et\thinspace al.\ \/}}

\def\eg{{e.g.\ \/}}

\def\lsim{~\rlap{$<$}{\lower 1.0ex\hbox{$\sim$}}}
\def\gsim{~\rlap{$>$}{\lower 1.0ex\hbox{$\sim$}}}

\begin{document}

\twocolumn[
\title{Distant clusters of galaxies in the ESO Imaging Survey}
\author{Marco Scodeggio$^1$, Lisbeth F. Olsen$^{1,2}$, Luiz N. da Costa$^1$\\
{\it $^1$ESO, K. Schwarzschild-Str. 2, D--85748 Garching b. M{\"u}nchen,
Germany}\\
{\it $^2$Astronom. Obs., Juliane Maries Vej 30, DK-2100 Copenhagen, 
Denmark}}
\vspace*{16pt}   

ABSTRACT.\ 
The results of a search for distant clusters of galaxies performed using
the I-band data obtained by the ESO Imaging Survey (EIS) are presented.
Cluster candidates are identified using a matched filter algorithm, that
provides not only an objective detection criterion, but also the means
to estimate the cluster redshift and richness.  A preliminary sample of
distant clusters has been obtained, containing 252 cluster candidates
with estimated redshift in the interval $0.2 \leq z\leq 1.3$ (median
redshift $z_{\rm med}\sim 0.4$) over an area of approximately 14 square
degrees.  The adopted selection criteria for the inclusion of cluster
candidates in this sample has been in general conservative, as the
primary concern has been the reliability of the candidates rather than
the completeness of the sample.
\endabstract]

\markboth{M. Scodeggio et al.}{EIS distant galaxy clusters}

\small

\section{Introduction}

One of the primary goals of the recently completed ESO Imaging Survey
(EIS; Renzini \& da Costa 1997) has been the preparation of a sample
of optically-selected clusters of galaxies over an extended redshift
baseline. Such a sample could be used for many different studies,
ranging from the evolution of the galaxy population in clusters, to
the evolution of the abundance of galaxy clusters, a powerful
discriminant of cosmological models. Until recently, only a handful of
clusters were known at redshifts $z > 0.5$; visual searches for high
redshift clusters were conducted by Gunn \etal (1986) and Couch \etal
(1991), but their samples are severely incomplete beyond $z \sim 0.5$;
at higher redshifts targeted observations in fields containing known
radio-galaxies and QSOs have produced a handful of cluster
identifications (\eg Dickinson 1995; Francis \etal 1996; Pascarelle
\etal 1996; Deltorn \etal 1997).  The first objective search for
distant clusters was conducted by Postman \etal (1996; hereafter P96)
using the 5-m telescope of the Palomar Observatory. In that survey 10
of the 79 cluster candidates that were identified have estimated
redshift $\geq 0.7$.  Further evidence for the existence of clusters
at high redshift has been obtained from X-ray (\eg Gioia \& Luppino
1994; Henry \etal 1997; Rosati \etal 1998), optical (\eg Connolly
\etal 1996; Zaritsky \etal 1997) and infrared (Stanford \etal 1997)
searches.  However, the existing samples are still small, and their
selection effects largely unknown.

In this paper we briefly describe a large sample of optically-selected
cluster candidates, obtained applying a matched filter algorithm to
the catalogs of objects detected on single 150 sec. EIS V- and I-band
exposures. The sample includes 252 candidates, identified over an area
of 14.4 deg$^2$. Further details on this work can be found in Olsen
\etal (1998a,b; hereafter O98a,b), and in Scodeggio \etal (1998).

\section{Observations and Data Reduction} 

EIS observations were conducted using the EMMI camera on the ESO 3.5m
New Technology Telescope. Observations were carried out over four
pre-selected patches of the sky, spanning a wide range in right
ascension, and cover a total area of 17 deg$^2$ in I-band. Only small
sub-areas were covered also by V- and B-band observations (3.5 and 1.5
deg$^2$, respectively). For this reason we concentrate here our
attention mostly on the I-band data. EIS observations consist of a
sequence of 150 sec exposures; however each point of a patch is imaged
twice, for a total integration time of 300 sec, using two frames
shifted by half an EMMI-frame both in right ascension and declination.
The easiest way of visualizing the global geometry of this mosaic of
frames is to consider two independent sets of them, forming contiguous
grids (in the following referred to as odd and even frames),
superposed and shifted by half a frame both in right ascension and
declination.

The data reduction is carried out automatically through the EIS
pipeline, described in Nonino \etal (1998; hereafter N98).  This
pipeline produces both coadded images and fully corrected single
frames, using the astrometric and photometric solution derived from
the global data reduction process.  The astrometric solution is found
relative to the USNO-A1 catalog.  The internal accuracy of the
astrometric solution is better than 0.03 arcsec, although the absolute
calibration suffers from the random and systematic errors of the
reference catalog. The photometric calibration is done in a two step
procedure, first bringing all the frames to a common photometric
zero-point, taking advantage of the overlap between the frames, then
making an absolute calibration based on external data.  The internal
accuracy of the photometric calibration is $\lsim 0.005$ ~mag. The
current absolute calibration uncertainty is $\lsim 0.2$ ~mag.  Further
details can be found in N98.

Among the various object catalogs produced by the EIS pipeline, we
use in this work the so-called odd and even catalogs, which are
single entry catalogs listing all objects detected in the odd or even
frames.  The reliability and completeness of these catalogs are
discussed in N98. Based on that analysis, it was estimated that the
single-frame odd and even I-band catalogs are $\sim$95\% complete to
$I=23.0$; with a differential completeness at this magnitude of 80\%
(the V-band ones have similar completeness to $V=24.0$). At this
limiting magnitude the contamination from spurious objects is
estimated to be approximately 20\%, with total contamination of the
catalog of $\sim$5\%.  The object classification was shown to be
reliable to $I \approx 21$. Below that limit the object classification
is not reliable any more, and all detected objects fainter than $I=21$
are taken to be galaxies. Already at this magnitude the fraction of
stars is found to be $\sim$25\% of the total number of objects, and
taking into account the steep rise of the galaxy number counts
faintward than $I = 21$, the contamination of the galaxy catalog by
stars can be considered negligible.

\section{Cluster detection}

Candidate galaxy clusters are objectively identified using a procedure
based on the matched filter algorithm described by P96. The details of
this EIS cluster finding pipeline are discussed by O98a, and only a
brief summary is given here.  The filter is derived from an
approximate maximum likelihood estimator, obtained from a model of the
spatial and luminosity distribution of galaxies within a cluster. This
distribution, as a function of magnitude $m$ and radial distance from
the cluster center $r_{\rm c}$, is represented as
\begin{equation} \label{eq:gal_distribution} 
C(r,m) = \Lambda_{\rm cl} P(r/r_{\rm c}) \phi(m-m^*) 
\end{equation} 
where $P$ is the cluster projected radial profile, $\phi$ is the
cluster luminosity function, and $\Lambda_{\rm cl}$ measures the
cluster richness.  The parameters $m^*$ and $r_{\rm c}$ are the
apparent magnitude corresponding to the characteristic luminosity of
the cluster galaxies, and the projected value of the cluster
characteristic scale length. From this model one can represent the
observed galaxy distribution $D$ as $D(r,m) = b(m) + C(r,m)$ where
$b(m)$ is the background number counts, and write an approximate
likelihood ${\cal L}$ of having a cluster at a given position as
\begin{equation}
\ln {\cal L} \sim \int {{C(r,m)}\over{b(m)}} D(r,m)~d^2r~dm
\end{equation} 

The functional dependence of ${\cal L}$ on the redshift-dependent
parameters $m^*$ and $r_c$ provides the mean of estimating the cluster
candidate redshift. By evaluating the function ${\cal L}$ for each
element of a two-dimensional array a filtered image of the input
galaxy catalog is created (the ``Likelihood map''). This procedure is
repeated a number of times, with the parameters $m^*$ and $r_c$ tuned
for different cluster redshift values.  Significant peaks in the
likelihood maps are identified independently in each map, using
SExtractor. A minimum size criterion is adopted, whereby only peaks
where more than $N_{\rm min}$ pixels have values above the detection
threshold $\sigma_{\rm det}$ are considered as potential
detections. The value of $N_{\rm min}$ is set to correspond to the
area of a circle with radius $1 r_{\rm c}$, while the value of
$\sigma_{\rm det}$ is kept constant at 2.  These parameters were
optimized using simulations described in O98a.

The lists of peaks identified in the various Likelihood maps are then
compared, and peaks detected at more than one filter redshift are
associated on the basis of positional coincidence. From this
association, likelihood versus $z$ curves are created, and the
redshift and richness estimates for each candidate are derived
locating the peak of the corresponding likelihood versus $z$
curve. Two richness parameters are derived for each candidate,
following P96. The first is obtained from the matched filter procedure
itself, using the parameter $\Lambda_{\rm cl}$, while a second
independent richness estimate is obtained to reproduce more closely
the conventional Abell richness parameter. The final cluster
candidates sample is then composed of those objects that persist for
at least three filter redshift tunings, and that have a richness
estimate $\Lambda_{\rm cl} \geq 30$. Simulations have shown these to
be quite conservative selection criteria, as noise peaks are most
likely to be associated with small values of the inferred richness,
and to be detected only for one (or few) particular redshift tuning of
the matched filter (see O98a for details).

\section{Results}

\begin{figure*}
\centering\mbox{\psfig{figure=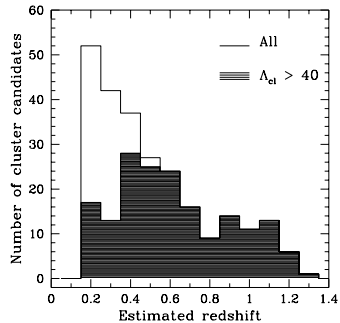,height=5.5cm}}
\caption[]{The estimated redshift distribution for the total sample 
of 252 EIS cluster candidates, and for the subset of 177 candidates
with richness estimate $>$ 40 (shaded portion of the histogram).}
\label{fig:z_all}
\end{figure*}

\begin{figure*}
\centering\mbox{\psfig{figure=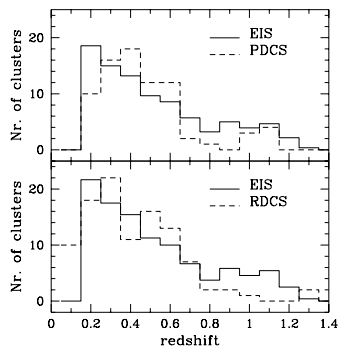,height=5.5cm}}
\caption[]{Comparison of the estimated redshift distributions for the
total EIS sample and for the sample of Postman \etal (upper panel);
and comparison between the estimated redshift distribution of the EIS
sample and the measured redshift distribution of the Rosati \etal
sample (lower panel). The EIS sample is scaled down to the same area
covered by the Postman \etal survey, and to the same total number of
clusters as in the Rosati \etal survey.}
\label{fig:z_comp}
\end{figure*}

The total sample of EIS cluster candidates, detected in I-band over
the four EIS patches, consists of 252 objects with estimated redshift
in the range $0.2 \leq z \leq 1.3$. The total area over which the
search was carried out is of 14.4 deg$^2$, and therefore the inferred
density of candidates is of 17.5 deg$^{-2}$. This value is slightly
larger than that found by P96, and could result from the somewhat
deeper magnitude limit achieved by EIS, with respect to the P96
data. The distribution of estimated redshifts for this total sample is
shown in Figure~\ref{fig:z_all}.  The median estimated redshift is
$z\sim 0.4$, and coincides with the value for the P96 sample. Note
however that the two redshift distributions differ somewhat, with the
EIS sample showing a more extended tail beyond $z \sim 0.6$ (upper
panel of Figure~\ref{fig:z_comp}). A similar comparison can be carried
out also with the sample of X-ray selected clusters of Rosati \etal
(1998). Using the cluster confirmations and redshift measurements
obtained up to December 1997 for that sample (Rosati, private
communication), we can see once again a general agreement (the median
redshift of the clusters in the Rosati \etal sample is also $z \sim
0.4$), and a more extended tail of high redshift objects in the EIS
sample (lower panel of Figure~\ref{fig:z_comp}). Spectroscopic
observations are now necessary to determine what fraction of EIS
cluster candidates are real objects, and to determine more accurately
their redshift.

For applications of cosmological interest, it is also interesting to
consider the distribution of rich (and presumably most massive)
clusters. This is presented in Figure~\ref{fig:z_all} as the shaded
portion of the histogram. From this distribution it is possible to deduce 
that the comoving density of cluster candidates is changing only very 
mildly as a function of redshift.

Over an area of 2 square deg. the search for cluster candidates has
been performed also using V-band data, independently from the search
in I-band. Afterwards the V and I-band candidate lists were compared,
and V-band detections associated to I-band ones. Of the 19 candidates
with estimated redshift $z \le 0.5$, 17 ($\sim$ 90\%) are also
detected in V-band; for those with estimated redshift $z > 0.5$, only
4 out of 16 are detected in V-band.  This result is not surprising
since the V-band data are relatively shallower the I-band ones. Using
the rule of thumb that the data should reach at least one magnitude
fainter than $m^*$ at the redshift of a given cluster to allow for its
detection, one can translate the galaxy catalog limiting magnitudes
adopted in the cluster search into limiting redshifts for cluster
detection. The I-band limit of $I=23.0$ then translates into a
limiting redshift between $\simeq 1.0$ (no-evolution model) and
$\simeq 1.3$ (passive evolution model), whereas the V-band limit of
$V=24.0$ translates into a limiting redshift between $\simeq 0.55$
(no-evolution) and $\simeq 0.7$ (passive evolution). 

\begin{figure*}
\centering\mbox{\psfig{figure=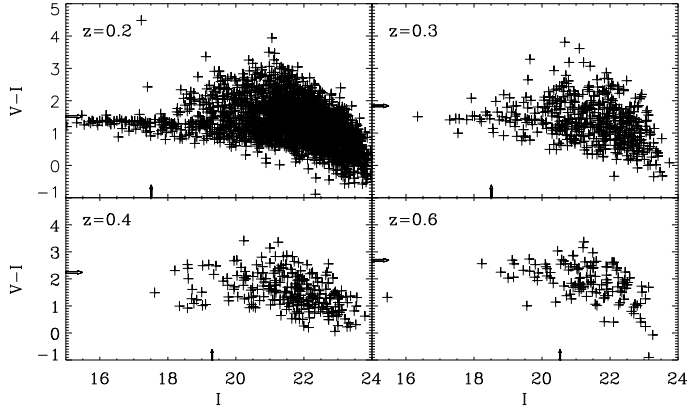,height=6cm}}
\caption[]{Color-magnitude diagrams observed for four cluster candidates
with estimated redshifts in range of $z=0.2$ to $z=0.6$, as indicated
in each panel. Also shown are the values of $m_I^*$ and the expected
colors of typical ellipticals at these redshifts.}
\label{fig:cm_diagram}
\end{figure*}

The availability of data in two passbands provides also an alternative
way of confirming cluster candidates and their estimated redshifts,
based on the detection of the sequence of cluster early-type galaxies
in a Color-Magnitude (CM) diagram.  In order to investigate this
possibility, a V-I vs I CM diagram was produced for each cluster
candidate, showing all galaxies within a radius of 0.75 h$^{-1}$ Mpc
(H$_0$ = 75 km s$^{-1}$/Mpc) from the nominal cluster center.
Fig.~\ref{fig:cm_diagram} shows four examples of such diagrams,
illustrating cases with estimated redshift in the range $0.2 \leq z
\leq 0.6$. At low redshift, the sequence of early-type galaxies is
clearly visible, but at $z \gsim 0.5$ the evidence for a CM relation
is, in most cases, less compelling.  In total, out of 35 cluster
candidates in the region of overlap of the V- and I-band images, there
are 19 with evidence for a CM relation, with estimated redshifts
extending out to $z \lsim 0.6$. Furthermore, the redshift estimates
based on the color and on the matched filter seem to agree, in most
cases, within 0.1. However, there are at least four cases where there
is a strong suggestion that the matched filter has overestimated the
redshift.

\section*{Acknowledgments}

Our most sincere thanks to the whole EIS team, for the great effort and
dedication shown in carrying out the survey over a very tight
timetable, without sacrifice of its overall quality and value.


\end{document}